# A Relative Study of Task Scheduling Algorithms in Cloud Computing Environment

Syed Arshad Ali, Student Member, IEEE, Mansaf Alam, Member, IEEE
Department of Computer Science
Jamia Millia Islamia
New Delhi, India
arshad158931@st.jmi.ac.in, malam2@jmi.ac.in

*Abstract*— Cloud Computing is a paradigm of both parallel processing and distributed computing. It offers computing facilities as a utility service in pay as par use manner. Virtualization, self-service provisioning, elasticity and pay per use are the key features of Cloud Computing. It provides different types of resources over the Internet to perform user submitted tasks. In cloud environment, huge number of tasks are executed simultaneously, an effective Task Scheduling is required to gain better performance of the cloud system. Various Cloud-based Task Scheduling algorithms are available that schedule the user's task to resources for execution. Due to the novelty of Cloud Computing, traditional scheduling algorithms cannot satisfy the cloud's needs, the researchers are trying to modify traditional algorithms that can fulfil the cloud requirements like rapid elasticity, resource pooling and on-demand self-service. In this paper the current state of Task Scheduling algorithms has been discussed and compared on the basis of various scheduling parameters like execution time, throughput, makespan, resource utilization, quality of service, energy consumption, response time and cost.

## I. INTRODUCTION

Cloud Computing is increasingly more attracting each industrial and educational communities. During the last decade, advancements in virtualization technology and service computing have empowered the value-effective comprehension of massive-scale statistics facilities that run massive portion of now a day's Internet programs and backend processing. The Cloud is a collection of computers or servers that can be interconnected collectively to offer resources to the customers. Cloud Computing joins a number of standards of grid, distributed and parallel computing. Cloud Computing is really getting access to resources and offerings needed functions in a dynamically changing environment. The user has no need to keep his own hardware and software resources, he can request application and computing services from the cloud. Advanced virtualization of resources helps the cloud to manage and maintain itself without user interference. According to the official National Institute for Standard Technology (NIST) definition [1], "Cloud Computing is a model for enabling ubiquitous, convenient, on-demand network access to a shared pool of configurable computing resources (networks, servers, storage, applications and services) that can be rapidly provisioned and released with minimal management effort or service provider interaction".

On-demand services are provided by the Cloud Computing on pay per use basis, the three popular service layers of Cloud Computing are: SaaS (Software as a Service), PaaS (Platform as a Service) and IaaS (Infrastructure as a Service) that helps academic institutions, public and private organizations to reduce their operational expenditures [2]. The fig. 1 illustrates the business model of Cloud Computing.

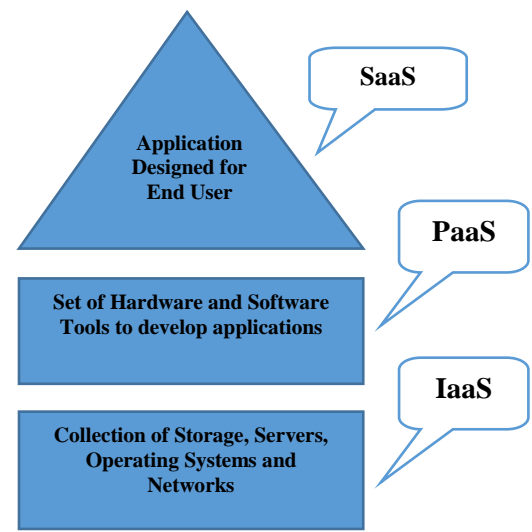

Fig. 1. Business Model of Cloud Computing

Cloud web hosting deployment models constitute the exact class of cloud environment and are mainly distinguished by means of the proprietorship, size and get admission to. The cloud deployment models come in four types: Private cloud, Public cloud, Community cloud and Hybrid cloud. Public cloud is a type of cloud web hosting wherein the cloud resources are delivered over Internet which is accessible for public usage. In Private cloud, infrastructure and services are provided over the private network that is shielded by a firewall which is under the governance of the IT department that belongs to the particular corporate. Hybrid cloud is an integration of multiple cloud hosting with multiple providers. Community cloud is a type of cloud hosting in which the setup is mutually shared between many organizations that belong to a particular community.

Cloud Computing has numerous characteristics which give benefits to the end user, optimistic features of a cloud are essential to permit services that definitely constitute the Cloud Computing model and satisfy expectations of consumers. The main features of Cloud Computing are self-serviced, per-usage metering and billing, elasticity and customization. For these features resource management plays a very important role. Management of resources is the method of assigning storage, energy, computing, and network resources to the user, for meeting target performance of the programs, cloud providers and customers of the cloud simultaneously. Virtualization is a technique of multiplexing cloud resources across applications and customers to provide efficient use of resources. According to study done in [3], the resource allocation process classifies into eight functional areas:

- Global scheduling of cloud resources.
- Resource request outlining.
- Resource utilization approximation.
- Application scaling and provisioning.
- Native scheduling of cloud resources.
- Resource pricing and profit maximization.
- Workload organization and
- Cloud managing system.

The main focus of this paper is on scheduling of cloud resources. The users request for the resources on demand, and the cloud provider is accountable for allocation of required resources to the user to avoid the violation of Service Level Agreement (SLA). The process of Task Scheduling instructs the scheduler to get tasks from the users and asks the cloud information service (CIS) for available resources and their properties. According to the availability of resources and Task Scheduling algorithm, scheduler schedules user submitted jobs on various resources as per requirements. Cloud scheduler is responsible to schedule multiple virtual machines (VMs) to different tasks.

The scheduling algorithm should be capable enough to handle the problems related to the resource allocation like resource contention, scarcity of resources, over provisioning of resources and resource fragmentation. When the number of tasks to be executed are large then the scheduling becomes difficult, therefore there is a need of an efficient scheduling algorithm. The fig. 2 illustrate the scheduling model in Cloud Computing environment. This paper is focusing on the various scheduling algorithms for cloud environment and compares them on the basis of various related parameters and find merits and demerits of these algorithms.

Following sections of the paper is presented as follows. In Section II various Task Scheduling parameters are discussed. Section III gives a brief analysis and comparison of different Task Scheduling algorithms in Cloud Computing environment. On the basis of analysis done, various issues and future directions are presented in Section IV and Section V concludes the paper.

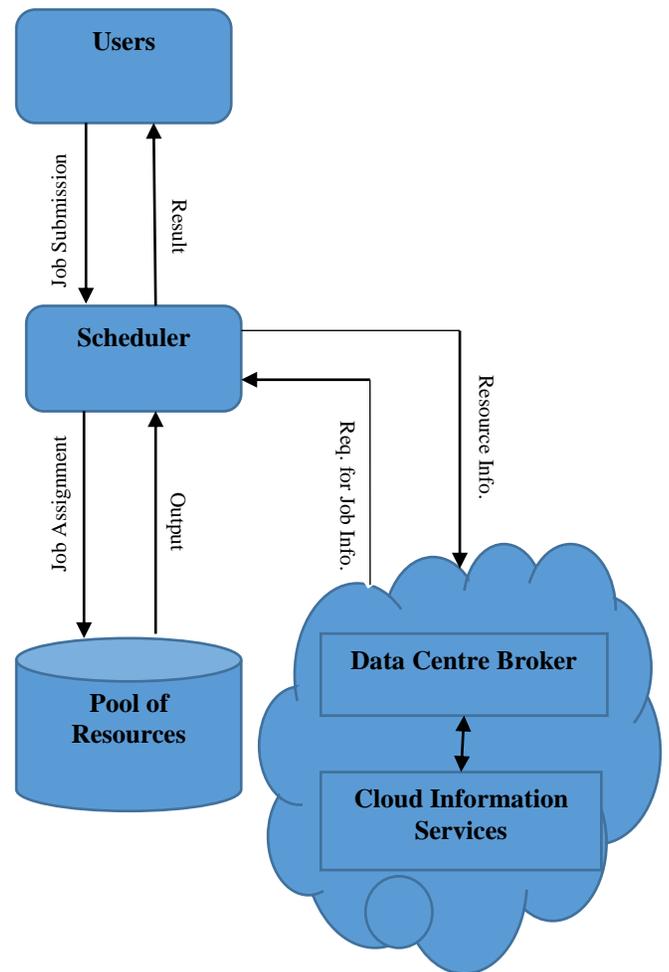

Fig. 2 Scheduling Model in Cloud Computing Environment

## II. TASKS SCHEDULING PARAMETERS

Several scheduling parameters are considered in different scheduling algorithms [4]. In this section, we will discuss these parameters for comparison of Task Scheduling algorithms.

1. Execution Time

The CPU time or burst time spent by the computer system for execution of a task is known as execution time, including the time consumed to provide system services for task execution.

2. Response Time

The amount of time taken by the system to reply the user task very first time for required service. That service may also be something from a memory fetch, to a disk IO, to an elaborate database question, or loading a full web page. Response time of the system should be minimum.

3. Makespan

The amount of time, from start to finish for completing a set of tasks. The makespan is the maximum time to complete all jobs. For better performance of the system makespan of scheduler must be minimum.

4. Throughput

Number of processes completed per unit time by the system is used to measure the performance of any system. In general throughput is the rate at which something can be processed. The main concern of scheduling algorithms is to increase the throughput of the system.

5. Resource Utilization

In addition to response time and throughput, another parameter for performance measurement of a system is resource consumption. How much amount of system resources are busy? is track using resource utilization. Scheduling algorithm should increase the utilization of resources.

6. Load Balancing

In computing, load balancing improves the distribution of workloads across more than one computing resources, a computer cluster, network hyperlinks, or disk drives. The reliability of system and concurrent user capacity are improved by a load balancer, which distributes the workload to various servers and reduces the overall load to be found on alone server.

7. Fault Tolerance

Fault tolerance is the property that allows for a procedure to continue running effectively within the incident of the failure in its components. It is an important parameter to check the capability of any system.

8. Energy Consumption

Energy consumption is the amount of resource energy used to produce the output. Energy consumption should be minimal. The Cloud Computing environments require a huge quantity of information and communication technologies gear reminiscent of servers, storage devices, communique network instruments and patron terminals. Hence, it's clear that the standard use of Cloud Computing offerings will broadly make contributions to a fast increase on ICT vigor consumption.

9. Scalability

It is a characteristic of a system, model or function that describes its ability to manage and participate in below multiplied or increasing workload. A process that scales well might be competent to hold or even broaden its level of efficiency when tested by using higher operational needs.

10. Performance

The accomplishment of a given task measured against pre-set identified requirements of completeness, cost, accuracy, and velocity. In computing performance is measured by the time and cost, a system should complete a user task in less time and minimum cost of services. A performance should be considered at both sides user and provider by scheduling algorithm.

11. Quality of Service

Best of provider considers user involvement restrictions like meeting cut-off date, efficiency, execution price, make span, and so forth. Everything are outlined in Service Level Agreement (SLAs) which is a contract file outlined between the cloud user and cloud service provider.

## III. VARIOUS TASK SCHEDULING ALGORITHMS

In this section, various Task Scheduling algorithms are discussed and related scheduling parameters are highlighted as pros and cons. Table -1 presents overall analysis and comparison of these Task Scheduling algorithms on the basis of these parameters. It consolidates different scheduling algorithms and corresponding parameters that they have supported.

1. Multi-Objective Task Scheduling Algorithm

Most of the scheduling algorithms minimize the makespan without consideration of reduction in cost of services, these algorithms are also applied on single-cloud environment. The algorithm discussed in [5] considers both parameters, the makespan and cost of service in a multi-cloud environment. It also considers the average utilization of cloud resources. The Cloud Min-Min Scheduling considers only minimization of makespan and Profit Based Task Scheduling (PBTS) considers only cost of services, none of them consider multiple parameters. Experiment results shows that, proposed algorithm minimizes cost and makespan simultaneously and maximizes average utilization of resources.

## 2. Multilevel Priority-Based Task Scheduling

Previous priority-based Task Scheduling algorithms schedules tasks according to the priority based on the size of program. In this proposed scheduling algorithm [6], six sigma control charts are used to prioritize various independent tasks of workflow. The scheduler divides tasks into different levels and allocates the resources in step with the aid requirements of the various tasks levels. The results of experiment show that this method is powerful to manage multilevel task of workflows. This algorithm effectively reduces execution time and makespan. The proposed algorithm mainly emphasizes on independent tasks of workflow by determining only one parameter, but it can be stretched for dependent tasks.

## 3. Load Balancing Task Scheduling Algorithm

For reducing the complexity, most of the scheduling algorithms do not consider the load balancing, which may appear uneven load. These algorithms do not consider the relationship among task allocated and the node load. If a task is allocated to an overloaded node, then the system performance will drastically decrease. The proposed algorithm [7] is based on weighted random and feedback mechanism which assures that exceptional nodes would not be overburdened and for general performance nodes have the capacity to execute tasks. The algorithm can efficiently stability the weight and balances the load of the unequal and overloaded nodes among the other nodes in the Cloud Computing environment.

## 4. Particle Swarm Optimization based Task Scheduling

This algorithm [8] considers the requirements of multidimensional Quality of Service (QoS). The equality of resource allocation results is selected through Berger model. The algorithm makes changes in the Particle Swarm Optimization (PSO) algorithm through correcting its limits dynamically. Then, it endorses a quality of service based dynamic particle swarm optimization (QoS-DPSO) algorithm. The experiment results show that the user tasks can be effectively executed and the fairness in allocation of system resource can be achieved by this algorithm. The effects of experiment suggest that this method can efficiently resolve Task Scheduling problem in cloud computing environment but it does not consider execution cost and response time.

## 5. Energy-Efficiency based Task Scheduling Algorithm

There is a huge power consumption in cloud based data centers, which led to attention on the Task Scheduling to decrease the power consumption. In this paper [9], a dynamic voltage and frequency scaling (DVFS) enabled Energy-efficient Workflow Task Scheduling algorithm (DEWTS) has been proposed, with the aim of acquire more power drop on top of assures the quality of service (QoS). The algorithm divides the parallel tasks in workflows to the suitable CPU, and use them at proper time to decrease power consumption and obtain demanded performance. Experimental outcome shows that the algorithm decreases the power consumption by 46.5% in parallel tasks execution along with balances the performance of scheduling.

## 6. Cuckoo Optimization based Task Scheduling Algorithm

When a set of tasks arrive at the cloud, the system is supposed to reply to all of them because it manages to gain the shortest response time. In this paper [10], Author uses Cuckoo algorithm to carry out one of these controls. The cause of the proposed technique is attaining an order of processing units such that the time of responses to tasks is minimized. The Cuckoo algorithm uses the range of virtual machines and the variety of tasks, with the aid of examining numerous orders of these machines, the proposed approach allocates hosts to tasks in a right manner. Due to small calculation and input data which cause least dependency on input data, makes the Cuckoo algorithm more popular. Because of levy flight general and local search can be done simultaneously. This algorithm uses Cuckoo algorithm for load balance in cloud calculations which decrease processing time of the system. In this algorithm bird nest is set of compute units, the population of birds is in relation with compute units' order and each levy flight switches task from one machine to another. Simulation effects display that the usage of Cuckoo algorithm for the aim of attaining the best order of processing devices can cause progressed performance parameters.

## 7. Green Energy-Efficient based Task Scheduling Algorithm

This proposed green energy-efficient scheduling algorithm [11] is based on priority Task Scheduling algorithm. The weight of virtual machines and the Service Level Agreement (SLA) required by user are used as parameters for priority to select VMs for executing tasks. This approach ensures the minimum requirement of resources and limits the over resource usage by a task. The power consumption of the server can be reducing when it is in idle state or low workload through Dynamic Voltage Frequency Scaling (DVFS) technique. Hence, this method can decrease the power consumption of servers and enhance the resource utilization. The simulation shows that the power dissipation in this method can decrease by 23% with compare to the method referenced in [12].

## 8. Fault-Tolerant Workflow Scheduling Algorithm

The proposed scheduling algorithm [13] select tasks for execution on the basis of two estimating models on-demand and spot instances to cut down service cost and at the same time fulfil workload deadline. The approach is vigorous counter to the performance variation of resources and also fault tolerant counter to the untimely death of spot instances. The algorithm uses just-in-time (JIT) and adaptive scheduling heuristic. The scheduler calculates the slack time for each ready job, the difference between the critical path time and deadline time. As job's slack time declines due to failure or deviation in performance of the system, this method adaptively shifts dynamically user jobs to on-demand instances. The experimental outcomes illustrate that these heuristic lessens execution time by 70% as compared to only one instance usage.

9. Adaptive Energy-Efficient Task Scheduling Algorithm

The traditional energy-efficient based Task Scheduling algorithms focus on a threshold to stable system performance and power consumption. But system parameters and application requirement cannot flexibly be adapted by the random threshold which makes the scheduling outcomes instable. In this method [14], author propose a two-phase Adaptive Energy-efficient Scheduling (AES), which incorporates the adaptive task replication policy with the Dynamic Voltage Scaling (DVS) technique. First of all, it proposed an adaptive threshold-enabled job replication strategy, which set ideal threshold, and secondly when any task has slack time then it selects the group of tasks on DVS-aided machines that can decrease voltage to decrease system power consumption. The experimental outcomes show that this method can save power consumption and increase system performance.

10. Online Optimization for Preemptable Task Scheduling

In this algorithm [15], author proposed a dynamic scheduling algorithm based on resource optimization technique for preemptable tasks in multi-cloud environment. They proposed two dynamic scheduling algorithms, Dynamic Cloud Min-Min Scheduling (DCMMS) and Dynamic Cloud List Scheduling (DCLS), for resource allocation mechanism. The original algorithms mentioned in [16] do not consider the task dependency. In these dynamic algorithms tasks dependencies are maintained by update the task set in each scheduling step. This dynamic method with updated knowledge of tasks provides substantial enhancement in the avoidance of resource contention situation. The power consumption can be reduced in cloud system by local mapping of resources with awareness of their energy consumption.

TABLE I. STUDY AND COMPARISON OF TASK SCHEDULING ALGORITHMS ON THE BASIS OF SCHEDULING PARAMETERS

| Scheduling Algorithm | Response Time | Execution Time | Throughput | Make Span | Resource Utilization | Energy Consumption | Load Balancing | Performance | Fault Tolerance | Scalability | Quality of Service | Cost |
|---|---|---|---|---|---|---|---|---|---|---|---|---|
| Multi-Objective Task Scheduling | - | - | - | ✓ | ✓ | - | - | - | - | - | ✓ | ✓ |
| Green Energy-Efficient based Task Scheduling Algorithm | - | - | - | - | ✓ | ✓ | - | - | - | - | ✓ | - |
| Multilevel Priority-Based Task Scheduling | - | ✓ | - | ✓ | - | - | - | - | - | - | - | - |
| Online Optimization for Preemptable Task Scheduling | - | ✓ | - | - | ✓ | ✓ | - | ✓ | - | - | - | - |
| Particle Swarm Optimization based Task Scheduling | - | - | - | - | ✓ | - | - | - | - | ✓ | ✓ | - |
| Load Balancing Task Scheduling Algorithm | - | - | - | - | ✓ | - | ✓ | ✓ | - | ✓ | - | - |
| Energy-Efficient based Task Scheduling Algorithm | - | - | - | - | ✓ | ✓ | - | ✓ | - | - | ✓ | - |
| Fault-Tolerant Workflow Scheduling Algorithm | - | ✓ | ✓ | - | - | - | - | ✓ | ✓ | - | - | - |
| Adaptive Energy-Efficient Task Scheduling Algorithm | - | - | - | - | ✓ | ✓ | - | ✓ | - | - | - | - |
| Cuckoo Optimization based Task Scheduling Algorithm | ✓ | - | - | - | - | - | ✓ | ✓ | - | - | - | - |

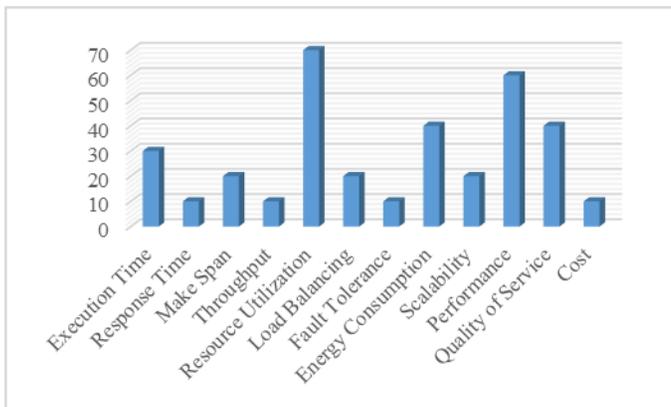

Fig. 3 Proportion of scheduling characteristics supported by various discussed Task Scheduling algorithms

## IV. ISSUES AND FUTURE DIRECTIONS

The study of different cloud based Task Scheduling algorithms shows that Task Scheduling plays an important part by providing scalability, on-demand resources, energy consumption, which enhanced the performance of the Cloud Computing system. Most of scheduling algorithms focused on the energy consumption and service cost, but there is still a lot of improvement is needed which constitutes the Cloud Computing towards the Green computing [17]. Most of the algorithms discussed here, perform scheduling on the basis of two or three scheduling parameters. An effective Task Scheduling algorithm can be design by adding more parameters to the existing algorithms which can enhance the system performance in near future.

## V. CONCLUSION

Task Scheduling is very important aspect for Cloud Computing. To provide scalability, resource pooling and on-demand self-service, an efficient Task Scheduling algorithm is required. In this paper, recently developed Task Scheduling algorithms in Cloud Computing environment have been studied and various Task Scheduling parameters are used to compare these algorithms. In which Particle Swarm Optimization and Cuckoo Optimization based Task Scheduling Algorithms are inspired from nature based algorithms and DVFS-enabled Energy-efficient Workflow, Green Energy-Efficient and Adaptive Energy-Efficient Task Scheduling Algorithms are based on energy consumption.

From fig. 3 we can see that resource utilization and system performance are covered in 60 to 70 percentage of discussed algorithms for improvement. Energy consumption and quality of service are improved in 40 percentage of algorithms. But there is still a lot of work required to improve fault tolerance and load balancing. An effective Task Scheduling algorithm should facilitate the cloud system to fulfill all the requirements made in the contract of Service Level Agreement (SLA) between the cloud providers and customer.

ACKNOWLEDGMENT

This work was supported by a grant from "Young Faculty Research Fellowship" under Visvesvaraya PhD Scheme for Electronics and IT, Department of Electronics & Information Technology (DeitY), Ministry of Communications & IT, Government of India.